\documentclass[onecollarge,natbib]{svjour2}
\bibpunct{[}{]}{,}{n}{}{,} 
\smartqed  
\usepackage{graphicx}
\usepackage{multirow}
\usepackage{subfigure}
\journalname{Few-Body Systems (APFB2011)}
\begin{document}

\title{\boldmath
Strange quarks and lattice QCD
}


\author{Anthony W.~Thomas         \and
        Phiala E.~Shanahan    \and
	Ross D.~Young
}


\institute{
          CSSM and CoEPP, School of Chemistry and Physics, \\
	University of Adelaide SA 5005, Australia
}

\date{Received: date / Accepted: date}

\maketitle

\begin{abstract}
The last few years have seen a dramatic improvement 
in our knowledge of the strange 
form factors of the nucleon. 
With regard to the vector from factors the level of 
agreement between theory and experiment gives us considerable confidence in our 
ability to calculate with non-perturbative QCD. The calculation of the 
strange scalar form factor has moved significantly 
in the last two years, with the 
application of new techniques which yield values considerably smaller than 
believed for the past 20 years. These new values turn out to have important 
consequences for the detection of neutralinos, 
a favourite dark matter candidate. 
Finally, very recent lattice studies have resurrected interest in the famed 
H-dibaryon, with modern chiral extrapolation of lattice data 
suggesting that  it 
may be only slightly unbound. We review some of the major 
sources of uncertainty in that 
chiral extrapolation.
\end{abstract}

\section{Introduction}
\label{intro}
One of the major initial successes of QED was the successful explanation of 
the famous Lamb shift in terms of the effect of vacuum polarization. For QCD 
the strange form factors of the nucleon 
occupy a position of comparable importance. 
While lattice QCD has accurately described a number of 
valence quark dominated hadronic 
properties, the strange form factors of the nucleon 
can only arise through quantum 
fluctuations in which a strange-anti-strange pair briefly bubble into and out 
of existence. 

It is now more than 20 years since it was realized that parity 
violating electron scattering (PVES) could provide a third, 
independent constraint on 
the vector form factors of the nucleon, thus allowing one to solve for the 
strange vector matrix elements~\cite{Mckeown:1989ir}. 
{}Following a series of state-of-the-art 
measurements at MIT-Bates, Mainz and JLab (for a recent review see 
Ref.~\cite{Paschke}), as well as systematic study 
of the relevant radiative corrections~\cite{Blunden:2011rd} 
and a careful global analysis~\cite{Young:2006jc}, we now 
know that the strange magnetic form factor is at most a few percent of the 
proton magnetic form factor at low-$Q^2$ and the strange electric radius 
is also at most a few percent of the proton charge radius.

The length of time since the initial studies of nucleon vector form factors 
in lattice QCD, by Leinweber and collaborators, is also around 20 years. 
In combination with chiral extrapolation 
of lattice data using finite range regularisation, 
indirect techniques developed by 
Leinweber and Thomas~\cite{Leinweber:1999nf} led to 
a very precise determination of both the strange 
magntic moment of the proton and its strange charge 
radius~\cite{Leinweber:2004tc} some 15 years later. 
Although it initially seemed as though these calculations disagreed with the 
PVES data, it is now clear that the agreement is excellent. Furthermore, 
precise, direct calculations of these form factors from 
the Kentucky group in just 
the last two years~\cite{Doi:2009sq} agree very well with the 
results of the earlier indirect work 
and with the experimental results. 

At present, the accuracy of the theoretical calculations exceeds that of the 
best experiments by almost an order of magnitude -- a remarkable exception in 
strong interaction physics. Clearly the challenge is there for a clever new 
idea to take us beyond the current experimental limitations. Nevertheless, 
this future challenge should not blind us to the tremendous achievements 
thus far and especially to the fact that QCD has passed its equivalent of 
the ``Lamb shift test'' with flying colours.

In the next section we briefly review the state of play with respect to the 
strange sigma commutator (the strange scalar form factor), for which the best 
value has taken a major shift in the last two or three years. We also point 
out the relevance to dark matter searches. In the third section we recall the 
recent surprises concerning the H-dibaryon 
and after  recalling the best current 
estimate of its mass we explore some of the potential uncertainties.

\section{Strange scalar form factor}
\label{sec:1}
In many ways it is fortunate that we have not had sufficient computing power 
to directly calculate hadron properties at the physical quark masses. Instead, 
we have been given data on a number of hadronic properties {\it with QCD} as 
a function of the masses of the quarks. 
This is of course information that Nature 
cannot give us but which is nevertheless 
an invaluable guide to how QCD actually 
works~\cite{Detmold:2001hq}. From those studies it is clear that for 
whatever reason (and we suggest one 
below) the properties of baryons and non-Goldstone mesons made of light quarks 
behave exactly as one would expect in a constituent 
quark model once the pion mass 
is above about 0.4 GeV. That is, all the famous, rapid, non-analytic variation 
associated with Goldstone boson loops is seen to 
disappear in this region. There 
has been some speculation that this scale (which corresponds to a quark mass 
around 40 MeV) may have something to do with 
the size of an instanton. However, it 
is clear that such behaviour does emerge maturally 
if one takes into account the 
finite size of such hadrons, which necessarily suppresses meson loops at large 
momentum transfer and at large mass for the Goldstone bosons. Put simply, 
meson loops are suppressed when the corresponding 
Compton wavelength is smaller than 
the size of the hadron emitting or absorbing the meson. 
Given that a typical hadron 
size is 1fm, and the Compton wavelength of a meson of 
mass 0.4 GeV only 0.5fm, one 
has a natural explanation.

This qualitative lesson from QCD itself, leads us to anticipate 
that the contribution 
of strange quarks to nucleon properties should be suppressed, 
as the mass of the kaon 
is 0.5 GeV  -- above the critical scale. 

This idea has also been developed to allow quantitative advances in the 
calculation of hadron properties using finite range regularization 
(FRR)~\cite{Young:2002ib}. This 
technique allows one to effectively resum the 
chiral expansion of hadronic properties 
and to accurately describe the variation of properties such as the mass of the 
baryon octet over a much larger range of quark masses 
than expected within naive 
chiral perturbation theory. 

Once one has an accurate parametrization of the 
mass of the nucleon as a function 
of pion and kaon mass, based on a fit to modern lattice data for the nucleon 
octet using FRR, it is trivial to extract the light quark sigma commutators 
by differentiation, using the Feynman-Hellmann theorem. It is this approach 
which has recently established that $\sigma_s$ is almost an order of magnitude 
smaller than had been generally believed for 20 years~\cite{Young:2009zb}. 
Very similar conclusions have been reached by several other modern lattice 
simulations~\cite{Ohki:2008ff,Toussaint:2009pz,Ohki:2009mt}, 
with everyone agreeing that the value is between 20 and 50 MeV. 
Although at first sight it is shocking that there can be such a large shift 
in a fundamental property of the nucleon, it is quite common with 
fundamental parameters that apparent convergence can be followed by a shift 
by far more than the quoted uncertainties when a new technique becomes 
available.

The importance of this new value for kaon condensation in dense matter 
remains to be investigated. It is in the search for dark matter that 
the new value of $\sigma_s$ has had its immediate 
impact~\cite{Giedt:2009mr}. In the 
minimal supersymmetric extension of the Standard Model, constrained by 
all particle physics and WMAP data, the so-called CMSSM, the favoured 
candidate for dark matter is the neutralino~\cite{Jungman:1995df}, 
a weakly interacting fermion 
with mass of order of a hundred GeV or more. With the old values of 
$\sigma_s$ its dominant interaction with the nucleon was through the strange 
quark -- which had led to Ellis and collaborators calling desperately for 
a more accurate determination of the strange quark sigma 
commutator~\cite{Ellis:2008hf}. 
The new value reported above not only reduces the expected cross section 
by an order of magnitude compared with earlier optimistic expectations but 
it also makes them far more accurate. We refer to the work of 
Giedt {\it et al.} for more details~\cite{Giedt:2009mr}.

\section{Uncertainties in the estimate of the mass of the $H$-dibaryon}
\label{sec:checks}
In this section we turn to the extremely exciting possibility that the 
$H$-dibaryon, which had been abandoned by many, may indeed be almost bound 
with respect to the $\Lambda - \Lambda$ threshold. This possibility 
was raised only in the last year by the NPLQCD and HAL 
collaborations~\cite{Beane:2010hg,Inoue:2010es}, 
which found the $H$ bound by tens of MeV at the unphysically large light 
quark masses where they could perform the calculations. The issue then becomes 
whether one can accurately extrapolate these results to the physical 
quark masses. Given our explanation of the effectiveness of FRR and the 
relatively large quark masses at which much of the data was taken, this 
seems the natural technique to apply. 
It is helpful that the corresponding chiral 
coefficient for the $H$, assuming that it is indeed a genuine multi-quark 
state and not a quasi-molecular state like the dueteron, were calculated 
in the early 80s by Mulders and Thomas~\cite{Mulders:1982da}.

By making a simultaneous fit to the precise octet baryon masses noted 
earlier, as well to the data for the $H$, Shanahan {\it et al.} were able 
to show that at the physical quark masses the $H$ is most likely slightly 
unbound: $m_H - 2 m_\Lambda = 13 \pm 14$ MeV~\cite{Shanahan:2011su}. 
In what follows 
we explore, in more detail than was possible in the original article, the 
uncertainties in this result. These turn out to be rather small. 

\subsection{Varation of the form of the regulator}
\label{subsec:varyreg}
We investigate the claim that varying the form of the regulator, $u(k)$,  
used with the FRR formalism does not change the results obtained for 
the binding of the $H$-dibaryon.
As well as the dipole form used for the primary fit, 
we investigate monopole and exponential forms:
\begin{eqnarray}
\textrm{monopole}: &\hspace{0.5cm} u_m(k) = \frac{\Lambda^2}{\Lambda^2+k^2}\\
\textrm{exponential}: & \hspace{0.5cm} u_e(k) = \textrm{exp}({\frac{-k^2}{\Lambda^2}}).
\end{eqnarray}
A sharp cut-off was also tested. Fits with each of these regulators 
yield $\chi^2$ values (per degree of freedom) between 0.48 and 0.49, 
with the $H$ unbound at the physical point by $13\pm14$~MeV 
(monopole and exponential forms), or $14\pm 16$~MeV (sharp cutoff). 
Octet masses from each of these fits match the results obtained from 
the fit with a dipole regulator to within the quoted precision. 
Best-fit parameter values for these fits are given in Table~\ref{tab:varyreg}. 
It is clear that the form of the regulator does not significantly 
effect the results, and the selection of a dipole form as 
regulator is justifiable.
\begin{table}[t]
\caption{Values of fit parameters for the octet and $H$-dibaryon data with monopole, exponential and sharp cutoff regulators used. Note that units are given in dimensionally appropriate powers of GeV.}
\begin{center}
\label{tab:varyreg}
\begin{tabular}{l l l l l l l l l}
\hline\noalign{\smallskip}
 \textbf{Monopole}& $\Lambda$ & $M^{(0)}$ & $\alpha$ & $\beta$& $\sigma$ & $B_0$ & $\sigma_B$ & $C_H$\\[3pt] \tableheadseprule\noalign{\smallskip}
best fit value & 0.63 & 0.87 & -1.66 & -1.16 & -0.49 & 0.019 & -2.27 & 5.56 \\  
error & 0.04 & 0.04 & 0.11 & 0.09 & 0.05 & 0.004 & 0.19 & 0.07 \\ \noalign{\smallskip}\hline\noalign{\smallskip}
\textbf{Exponential} & & & & & & & & \\[3pt] \tableheadseprule\noalign{\smallskip}
best fit value & 0.81 & 0.86 & -1.75 & -1.22 & -0.53 & 0.020 & -2.44 & 5.71 \\  
error & 0.05 & 0.04 & 0.11 & 0.10 & 0.05 & 0.004 & 0.21 & 0.10 \\ \noalign{\smallskip}\hline\noalign{\smallskip}
\textbf{Sharp cutoff} & & & & & & & & \\[3pt] \tableheadseprule\noalign{\smallskip}
best fit value & 0.57 & 0.83 & -1.94 & -1.36 & -0.61 & 0.018 & -2.81 & 5.95 \\  
error & 0.02 & 0.03 & 0.10 & 0.08 & 0.05 & 0.003 & 0.18 & 0.10 \\ \noalign{\smallskip}\hline
\end{tabular}
\end{center}
\end{table}

\subsection{Variation of form factor mass for the $H$ dibaryon}
\label{subsec:varyL}
In the primary fit, the regulator mass $\Lambda$ was taken to be the 
same for both the $\Lambda$ hyperon and the $H$-dibaryon. 
When the ratio $\frac{\Lambda_H}{\Lambda_\Lambda}$ is varied, however, 
we do find a significant variation in binding, with little associated 
variation in the quality of the fit. For $\Lambda_H$ $20 \%$ smaller 
than $\Lambda_\Lambda$, the $H$ is bound at the physical point 
by $2 \pm 14$~MeV. Note, however, that in this case the best-fit 
value of the chiral coefficient of the $H$, called $C_H$, grows to 
approximately 2.5 times the Mulders-Thomas estimate~\cite{Mulders:1982da}. 
{}For $\Lambda_H$ $20 \%$ larger than $\Lambda_\Lambda$, the $H$ is 
unbound by $23 \pm 13$~MeV. In this case, the best-fit value of $C_H$ 
in fact moves closer to the Mulders-Thomas estimate; to within $10 \%$. 
The $\chi^2$ per degree of freedom for these fits are 0.47 and 0.49, 
respectively, a comparable quality of fit to the primary (which 
had $\chi^2$/dof = 0.48). Best-fit values 
are shown in Table~\ref{tab:Ldiff}.

While naively one may expect that $\Lambda_H$ should be somewhat smaller 
than $\Lambda_\Lambda$, corresponding to the expectation that the $H$ 
should be larger than the $\Lambda$ baryon in coordinate space, 
this is obscured by the regularisation scheme. As a different choice 
of $\Lambda$ corresponds to a difference in the resummation of 
higher-order terms in the formal chiral expansion, it is likely that 
the dominant sensitivity of the fit to a variation in $\Lambda$ will 
vanish, being compensated for by variations of the fit parameters 
$B_0$ and $\sigma_B$.

{}For this reason, we suggest that, even  
though the inclusion of a mass parameter $\Lambda$ was motivated by 
the physical idea that loop processes with momenta greater than 
$\Lambda \sim R^{-1}$ are suppressed, in the regularized FRR formalism 
this parameter simply becomes a 
scale beyond which the effective theory is no longer valid. That is, 
the parameter $\Lambda$ no longer strictly characterizes a physical 
property of a model. We thus assert that 
$\Lambda_H = \Lambda_\Lambda = \Lambda$ is an appropriate choice.
\begin{table}[t]
\caption{Values of fit parameters for the octet and $H$-dibaryon data, 
with $\Lambda_H$ $20\%$ smaller or larger than $\Lambda_\Lambda$. 
All quantities are given in appropriate powers of GeV.}
\begin{center}
\label{tab:Ldiff}
\begin{tabular}{l l l l l l l l l l}
\hline\noalign{\smallskip}
& & $\Lambda$ & $M^{(0)}$ & $\alpha$ & $\beta$& $\sigma$ & $B_0$ & $\sigma_B$ & $C_H$\\[3pt]\tableheadseprule\noalign{\smallskip}
\multirow{2}{*}{\textbf{$\frac{\Lambda_H}{\Lambda_\Lambda}$=0.8}}&best fit value & 1.02 & 0.86 & -1.71 & -1.20 & -0.51 & 0.08 & -3.41 & 10.03 \\  
&error & 0.06 & 0.04 & 0.12 & 0.10 & 0.05 & 0.01 & 0.26 & 0.50 \\ \noalign{\smallskip}\hline\noalign{\smallskip} 

\multirow{2}{*}{\textbf{$\frac{\Lambda_H}{\Lambda_\Lambda}$=1.2}} & best fit value & 1.01 & 0.86 & -1.70 & -1.19 & -0.51 & -0.02 & -1.76 & 3.69 \\
& error & 0.06 & 0.04 & 0.11 & 0.09 & 0.05 & 0.01 & 0.15 & 0.05 \\ \noalign{\smallskip}\hline
\end{tabular}
\end{center}
\end{table}
%
%

\subsection{Variations of the phenomenological inputs}
\label{subsec:vpi}
Certain phenomenologically determined constants were input into the fit 
of the binding of the $H$-dibaryon. In particular, $f$, the meson decay 
constant in the chiral limit, and the baryon-baryon-meson coupling constants,
$F$ and $C$, were set to values determined by chiral perturbation theory 
and SU(6) symmetry.

We test the robustness of the method by varying these parameters 
by $\pm 10\%$, and noting that the dependence of the binding of the $H$ 
on these variations is small compared to the statistical error associated 
with our fit. We find the $H$-dibaryon unbound by $12$ to $14$~MeV, 
$9$ to $17$~MeV, and  $11$ to $15$~MeV at the physical point, for 
$10\%$ variations in $f$, $F$ and $C$ respectively. There is little 
associated variation in the $\chi^2$ values for these fits, 
all lying between 0.44 and 0.5 (compared to 0.48 for the primary fit). 
We also note that the values of the octet baryon masses from each of 
these fits are compatible, within quoted error, with those of the 
primary fit. Table~\ref{tab:varyparam} gives best-fit parameters for 
each of the fits considered.
%
\subfigtopskip=12pt
\begin{table}[t]
\caption{Best-fit parameters for the octet and $H$-dibaryon lattice data, 
setting $f$, $F$ and $C$ to be $10 \%$ smaller or larger than 
their phenomenological values, given in Ref.~\protect\cite{Shanahan:2011su}. 
All quantities are given in appropriate powers of GeV.}
\label{tab:varyparam}
\centering
\subtable[Vary $f$, the meson decay constant in the chiral limit.]{
\begin{tabular}{l l l l l l l l l}
\hline\noalign{\smallskip}
 \textbf{f $\rightarrow$ 0.9f}& $\Lambda$ & $M^{(0)}$ & $\alpha$ & $\beta$& $\sigma$ & $B_0$ & $\sigma_B$ & $C_H$\\ [3pt]\tableheadseprule\noalign{\smallskip}
best fit value & 0.92 & 0.85 & -1.86 & -1.28 & -0.58 & 0.015 & -2.40 & 7.03 \\  
error & 0.06 & 0.04 & 0.13 & 0.11 & 0.06 & 0.005 & 0.22 & 0.11 \\ \noalign{\smallskip}\hline\noalign{\smallskip}
\textbf{f $\rightarrow$ 1.1f} & & & & & & & & \\ [3pt]\tableheadseprule\noalign{\smallskip}
best fit value & 1.12 & 0.87 & -1.59 & -1.13 & -0.46 & 0.08 & -1.25 & 6.29 \\  
error & 0.14 & 0.04 & 0.11 & 0.09 & 0.05 & 0.02 & 0.14 & 0.02 \\ \noalign{\smallskip}\hline
\end{tabular}
\label{tab:f}
}
\subtable[Vary the baryon-baryon-meson coupling constant $F$.]{
\begin{tabular}{l l l l l l l l l}
\hline\noalign{\smallskip}
 \textbf{F $\rightarrow$ 0.9F}& $\Lambda$ & $M^{(0)}$ & $\alpha$ & $\beta$& $\sigma$ & $B_0$ & $\sigma_B$ & $C_H$\\ [3pt]\tableheadseprule\noalign{\smallskip}
Best fit value & 1.05 & 0.86 & -1.60 & -1.26 & -0.48 & 0.027 & -2.45 & 5.39 \\  
error & 0.12 & 0.03 & 0.11 & 0.10 & 0.05 & 0.002 & 0.21 & 0.08 \\ \noalign{\smallskip}\hline\noalign{\smallskip}
\textbf{F $\rightarrow$ 1.1F} & & & & & & & & \\[3pt]\tableheadseprule\noalign{\smallskip}
best fit value & 0.99 & 0.86 & -1.81 & -1.12 & -0.55 & 0.011 & -2.24 & 5.93 \\  
error & 0.12 & 0.04 & 0.13 & 0.09 & 0.06 & 0.005 & 0.19 & 0.09 \\ \noalign{\smallskip}\hline
\end{tabular}
\label{tab:F}
}
\subtable[Vary the baryon-baryon-meson coupling constant $C$.]{
\begin{tabular}{l l l l l l l l l}
\hline\noalign{\smallskip}
 \textbf{C $\rightarrow$ 0.9C}& $\Lambda$ & $M^{(0)}$ & $\alpha$ & $\beta$& $\sigma$ & $B_0$ & $\sigma_B$ & $C_H$\\  [3pt]\tableheadseprule\noalign{\smallskip}
best fit value & 1.06 & 0.86 & -1.71 & -1.14 & -0.51 & 0.013 & -2.24 & 5.91 \\  
error & 0.13 & 0.04 & 0.12 & 0.09 & 0.05 & 0.005 & 0.19 & 0.10 \\ \noalign{\smallskip}\hline\noalign{\smallskip}
\textbf{C $\rightarrow$ 1.1C} & & & & & & & & \\[3pt]\tableheadseprule\noalign{\smallskip}
best fit value & 0.98 & 0.86 & -1.71 & -1.25 & -0.51 & 0.025 & -2.50 & 5.41 \\  
error & 0.11 & 0.04 & 0.12 & 0.11 & 0.05 & 0.003 & 0.21 & 0.08 \\\noalign{\smallskip}\hline
\end{tabular}
\label{tab:C}
}
\end{table}

\subsection{Variation of the chiral coefficient $C_H$ }
\label{subsec:fixch}
\begin{figure}
\centering
\includegraphics[scale=0.95]{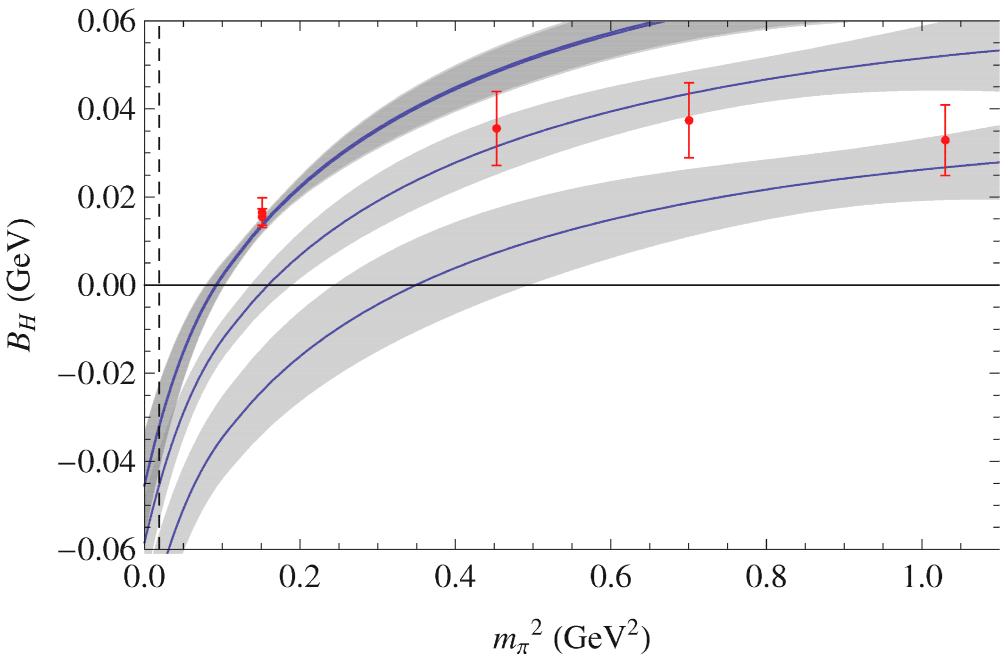}
\caption{Binding energy of the $H$-dibaryon versus pion mass 
squared, resulting from our chiral fit with $C_H$ fixed at the 
Mulders-Thomas value~\protect \cite{Mulders:1982da}, for several values 
of the kaon mass at which the simulations by HAL QCD and NPLQCD 
were carried out.}
\label{fig:bindingfixch}
\end{figure}

Figure~\ref{fig:bindingfixch} and Table~\ref{tab:bestfittabch} give the 
results of performing fits to the octet and binding data, while holding 
$C_H \sim 4.1$ fixed at the Mulders-Thomas value~\cite{Mulders:1982da},
calculated using SU(6) symmetry. As is 
clear from Figure~\ref{fig:bindingfixch}, the fit is not as good as the 
primary results. In fact, there is a $\chi^2$ of almost 2 per 
degree of freedom associated with the binding portion of the fit. 
At the physical point, the $H$-dibaryon is unbound by $30 \pm 9$~MeV.
\begin{table}[t]
\caption{Values of the fit parameters for the octet and $H$-dibaryon 
data corresponding to the fit shown in Figure~\ref{fig:bindingfixch}. 
Note that units are given in dimensionally appropriate powers of GeV.}
\label{tab:bestfittabch}
\begin{center}
\begin{tabular}{l l l l l l}
\hline\noalign{\smallskip}
 & $\Lambda$ & $M^{(0)}$ & $\alpha$ & $\beta$  \\[3pt]\tableheadseprule\noalign{\smallskip}
best fit value & 0.77 & 0.93 & -1.46 & -0.98   \\
error & 0.06 & 0.03 & 0.13 & 0.10   \\ \noalign{\smallskip}\hline\noalign{\smallskip}
& $\sigma$ & $B_0$ & $\sigma_B$ & \\ [3pt]\tableheadseprule\noalign{\smallskip}
& -0.40 & 0.06 & -1.20 &  \\
& 0.06 & 0.03 & 0.08 & \\ \noalign{\smallskip}\hline
\end{tabular}
\end{center}
\end{table}
%

%
%

\section{Conclusion}
\label{sec:4}
We have briefly reviewed the experimental and theoretical status of the 
strange vector form factors of the proton. Their importance is that since they
involve only ``disconnected diagrams'' the capacity of non-perturbative QCD 
to calculate them is comparable to that of the successful calculation of the 
Lamb shift in QED. As we showed, QCD passes this test with flying colours.

With regard to the strange scalar form factor, there has been a dramatic 
downward shift in the past couple of years, with the contribution to 
the mass of the nucleon arising through the strange quark mass 
now around 3\%, rather than 30\%. As we explained, this is critical to 
the interpretation of searches for neutralino dark matter.

Finally, we pointed out the very exciting new developments concerning 
the $H$-dibaryon, which now appears to be only slightly unbound. Although 
the calculation of Shanahan {\it et al.}~\cite{Shanahan:2011su}, which 
found the $H$ unbound by $13 \pm 14$ MeV assumed that it was a true 
multi-quark state, similar values have since been reported under quite 
different assumptions~\cite{Beane:2011iw,Haid}. This makes it urgent to 
pursue, initially at J-PARC, the sort of work reported in 
Ref.~\cite{Yoon:2007aq}, which has already given 
a hint of the existence of such a 
state.

\begin{acknowledgements}
This work was supported by the University of Adelaide and by the 
Australian Research Council through grants FL0992247 (AWT) and 
DP110101265 (RDY).
\end{acknowledgements}
%


\end{document}